\documentclass[aoas,MSNbibl,nameyear,dvips]{arximspdf}
\usepackage{graphicx}

\doi{10.1214/13-AOAS692} 
\volume{8}
\issue{1}
\pubyear{2014}
\firstpage{406}
\lastpage{429}

\makeatletter
\newcommand{\E}{\mathrm{E}}
\newtheorem{theorem}{Theorem}
\newproclaim{assumption}{Assumption}
\makeatother

\begin{document}
\begin{frontmatter}

\title{A functional data analysis approach for~genetic~association studies}
\runtitle{FDA approach for genetic association studies}

\begin{aug}
\author[A]{\fnms{Matthew} \snm{Reimherr}\corref{}\ead[label=e1]{mreimherr@psu.edu}\ead[label=u1,url]{www.personal.psu.edu/mlr36}}
\and
\author[B]{\fnms{Dan} \snm{Nicolae}\ead[label=e2]{nicolae@galton.uchicago.edu}\ead[label=u2,url]{www.stat.uchicago.edu/\textasciitilde nicolae}}
\runauthor{M. Reimherr and D. Nicolae}
\affiliation{Pennsylvania State University and University of Chicago}
\address[A]{Department of Statistics\\
Pennsylvania State University\\
State College, Pennsylvania 16801\\
USA\\
\printead{e1}\\
\printead{u1}} 
\address[B]{Departments of Statistics and Medicine\\
University of Chicago\\
Chicago, Illinois 60637\\
USA \\
\printead{e2}\\
\printead{u2}}
\end{aug}

\received{\smonth{5} \syear{2013}}
\revised{\smonth{8} \syear{2013}}

%
\begin{abstract}
We present a new method based on Functional Data Analysis (FDA) for
detecting associations between one or more scalar covariates and a
longitudinal response, while correcting for other variables. Our
methods exploit the temporal structure of longitudinal data in ways
that are otherwise difficult with a multivariate approach.
Our procedure, from an FDA perspective, is a departure from more
established methods in two key aspects. First, the raw longitudinal
phenotypes are assembled into functional trajectories prior to
analysis. Second, we explore an association test that is not directly
based on principal components. We instead focus on quantifying the
reduction in $L^2$ variability as a means of detecting associations.
%
%
Our procedure is motivated by longitudinal genome wide association
studies and, in particular, the childhood asthma management program
(CAMP) which explores the long term effects of daily asthma treatments.
We conduct a simulation study to better understand the advantages
(and/or disadvantages) of an FDA approach compared to a traditional
multivariate one. We then apply our methodology to data coming from
CAMP. We find a potentially new association with a SNP negatively
affecting lung function. Furthermore, this SNP seems to have an
interaction effect with one of the treatments.
\end{abstract}

%
\begin{keyword}
\kwd{Functional data analysis}
\kwd{longitudinal data analysis}
\kwd{genome wide association study}
\kwd{functional linear model}
\kwd{functional analysis of variance}
\kwd{hypothesis testing}
\end{keyword}

\end{frontmatter}

\section{Introduction}\label{sec1}

\subsection{The childhood asthma management program}

The childhood asthma management program, CAMP, is a multi-center,
longitudinal clinical trial designed to better understand the long term
impact of two common daily asthma medications, Budesonide and
Nedocromil, on children [The Childhood Asthma Management Program
Research Group (\citeyear{camp1999,camp2000})]. Subjects, ages
5--12, with asthma were selected, randomly assigned a particular
treatment (one of the two drugs or placebo) and monitored for several
years. At each clinical visit a number of measurements were taken, but
the primary one we focus on here is \textit{forced expiratory volume
in one second} or FEV1, which measures the development of the lungs.
The goal of the present paper is to associate FEV1, measured
longitudinally, with single nucleotide polymorphisms (SNPs) and to
detect possible SNP by treatment interactions, while correcting for
other covariates such as age and gender. Analyzing longitudinal data
can be challenging and often such measurements are converted to scalars
where univariate methods can readily be applied [\citet{tantisiraetal2011}]. However, since the subjects of this study are
children, the development of the child over time is also of great
interest. This development can be complicated and nonlinear as the
child ages, thus, a flexible framework to allow for such patterns is
desirable. Conversely, analyzing hundreds of thousands of SNPs requires
powerful procedures which exploit any structure inherent in the data.
It is reasonable to think that, while a child's lungs develop
``nonlinearly,'' their development is still relatively smooth over time,
and that major daily fluctuations in FEV1 are primarily noise
independent of the underlying development. Finally, while all children
make their clinical visits at approximately the same time, the spacing
between visits varies throughout the study. So a procedure which is
relatively robust against differences in temporal spacing would be vital.

%
%
\begin{figure}[b]
\includegraphics{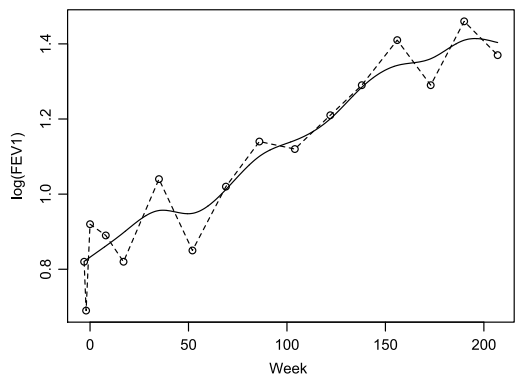}
\caption{Plot of one subject from the CAMP study. Circles represent
observed values which are linearly interpolated by the dashed line,
while the solid line is calculated using smoothing splines.}\label{ffdexample}
\end{figure}

\subsection{Functional data analysis}

Given the nature of the data and our goals, we develop and present a
framework for association testing based on functional data analysis.
Over the last two decades an abundance of high frequency data and
complex longitudinal studies have driven the development of inferential
statistical tools for samples of objects which can be viewed as
functions or trajectories. Tools falling under the umbrella of \textit
{functional data analysis} (FDA) have been applied to areas such as
human growth patterns [\citet{chenmuller2012,verzelenetal2012}],
gene expression [\citet{tangmuller2009}],
credit card transaction volumes [\citet{kokoszkareimherr2013}],
geomagnetic activity patterns [\citet{gromenkokokoszka2013}],
and neuroimaging [\citet{reissetal2011,crainiceanuetal2011}], to
name only a few. The driving view in FDA is that certain data
structures can be viewed as observations from a function space. To
illustrate this point, consider Figure~\ref{ffdexample}.
We plot the values of $\log(\mathrm{FEV}1)$ for one particular CAMP subject as
circles, with linear interpolation indicated by a dashed line. A
nonparametric smoother based on B-splines is also plotted as a solid
line. As we can see, the B-splines have generated a curve which has
smoothed out a lot of the inherent noise in the data, giving a clearer
picture of lung development in the child.

%
%
\begin{figure}
\includegraphics{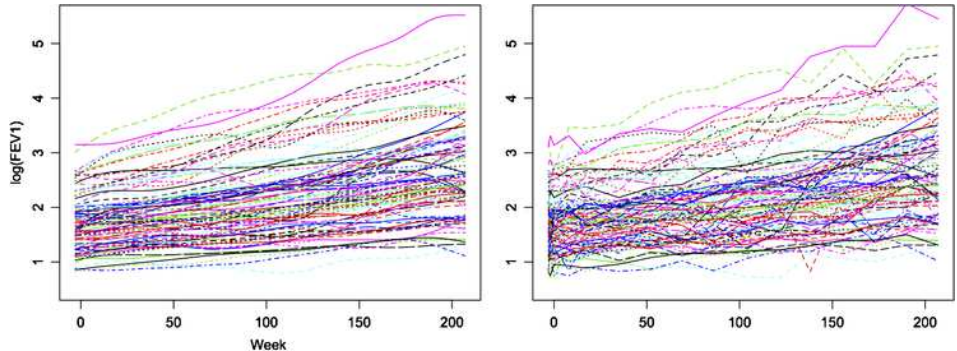}
\caption{Plots of the $\log(\mathrm{FEV}1)$ curves for the first one hundred
subjects. The left panel displays curves obtained via smoothing
splines, while the right panel displays curves which are linearly
interpolated.}\label{fall}
\end{figure}

We can obtain such a curve for every subject (and actually pool
information across children to help generate the curves). We plot the
resulting curves for the first 100 CAMP subjects (of 540 total) in
Figure~\ref{fall}, alongside their unsmoothed interpolated analogs.
We can see that the nonparametric smoothers have decreased the noise
inherent in the data, resulting in a clear smooth trajectory for each
child's development. Our goal is then to associate the patterns we see
in the curves with SNPs and to explore SNP by treatment interactions.
However, analyzing such objects is challenging because they are
inherently infinite-dimensional objects. The approach we outline here
exploits the assumption that the functions come from a~Hilbert space by
using a \textit{functional linear model} to relate the functional
response variables to the univariate covariates. Define $Y_n(t)$ to be
the value at time $t$ of the smoothed $\log(\mathrm{FEV}1)$ curve for the
$n$th subject. We then use the following linear model,
%
\begin{eqnarray}
Y_n(t) &=& \alpha(t) + \sum_{j=1}^J
\beta_{1,j}(t) X_{1;j,n}+ \sum_{k=1}^K
\beta_{2,k}(t) X_{2;k,n} + \varepsilon_n(t)
\label{eflm}
\end{eqnarray}
for $n=1, \ldots, N$, where $N$ is the total number of subjects, and
the total number of covariates is $J+K$. Here the $\{Y_n\}$, $\{
X_{1;j,n}\}$ and $\{X_{2;k,n} \}$ are observed. The first set of
covariates $\{X_{1;j,n}\}$ will include those we wish to correct for
(gender, age, etc.), while the second set will include those covariates
we wish to test the nullity of (SNPs, treatment effects, etc.).
Therefore, the goal of this paper is test the hypothesis
\[
H_0\dvtx  \beta_{2,k}(t) \equiv0\qquad\mbox{for }k = 1, \ldots, K.
\]
To test if $H_0$ is true, we propose using the reduction in the sum of
squared norms as a testing procedure. Our approach is a kind of
functional analog to the reduction in sum of squares used in univariate
ANOVA methods. This differs from current FDA approaches which are
usually based on principal component analysis and attempt to project
the infinite-dimensional problem down to a multivariate problem. As we
will note in Section~\ref{sapp}, FDA methods based on PCA seem to
have nontrivial stability problems which become especially evident when
carrying out hundreds of thousands of tests.

While the literature on FDA and functional linear models is quite
large, we provide some key references in terms of carrying out
hypothesis tests with a functional linear model.
\citet{cardotfms2003} examine a PCA-based testing procedure for
a scalar response/functional covariate model which \citet
{kokoszkamaslovasz2008} extends to the fully functional setting.
\citet{antoniadissapatinas2006} examine a mixed-effects model
for the modeling of functional data and utilize a wavelet decomposition
approach to carry out inference. \citet{zhangchen2007} analyze
the effect of the ``smoothing first then inference'' approach to
handling functional data (like the approach considered presently) and
consider an $L^2$ normed approach for testing the nullity of the
covariates in a functional linear model. Finally, \citet
{reissetal2010} combine a basis expansion approach for model fitting
and a permutation testing approach for local hypothesis testing in a
functional linear model which is implemented in the \texttt{R} package \texttt{Refund}.

\subsection{Alternative methodolgies}
The application of FDA methods in main stream human genetics research
is present but still in its infancy. \textit{Functional mapping}
methods [\citet{macawu2002,wulin2006}] in genetics are similar in
spirit to FDA methods, but, as of yet, have not taken advantage of the
full flexibility of FDA methods. In most functional mapping settings
very specific shapes for describing the data are instilled in the
models with fairly simple error structures, whereas an FDA approach
utilizes nonparametric methods to allow for greater flexibility. More
generally, the application of FDA methods to longitudinal data is a
very active area of research; see, for example,
\citet{fanzhang2000,yaomullerwang2005JASA,hallmullerwang2006},
to name only a few. Typically, FDA approaches for longitudinal data
focus on estimation and obtaining principal components which differs
with our goal of powerful hypothesis testing.

More traditional longitudinal approaches model the dependence between
observations from the same subject by introducing random effects. Such
an approach typically assumes a very specific structure for the
dependence between observations, but the question of how to carry out
the desired hypothesis tests is not solved by introducing mixed effects.
One could include fixed effects in the form of polynomial functions
over time, but given how the children are developing, it is difficult
to say which functions would be most appropriate. Another approach
would be to use time series to model the within subject dependence.
However, the time points are not evenly spaced, the data is not
stationary, and one would still have the problem of how to carry out
the hypothesis testing.

\subsection{Overview}
The remainder of the paper is organized as follows. In Section~\ref
{smeth} we present our approach based on the reduction of sum of
squared norms. We first present the single predictor case for
illustrative purposes and conclude with the general case that allows
one to correct for covariates, test the nullity of factor variables, or
test the nullity of multiple covariates. In Section~\ref{sframe} we
explore some of the major differences between our procedure and
previous FDA approaches. In particular, we show how the difference can
be thought of as a difference in weighting schemes on the scores. This
view allows for an entire family of testing procedures by choosing
different weights. In Section~\ref{ssims} we present a small-scale
simulation study. We show how the smoothness of the underlying
functions influences the power of our procedure as compared to a
PC-based procedure and a traditional multivariate one. In Section~\ref
{sapp} we apply our methods to a genome wide association study on
childhood asthma that has proven difficult to analyze using traditional
methods. We show that there may in fact be a gene by treatment
interaction, but further validation is needed. We conclude the paper
with discussion of our results in Section~\ref{sdisc}. All asymptotic
results are provided in Appendix \ref{sasymptotics}, while proofs are
given in Appendix \ref{sproofs}.

\section{Methodology}\label{smeth}
To construct the trajectories, we first apply a subject by subject
B-splines smoother. Some type of smoothing is useful
for the CAMP data due to the inherent noise in spirometer readings. A
splines-based smoother is especially useful given the fairly smooth and
nearly monotonic structure of the trajectories. The smoothing
parameter is chosen by leave-one-subject-out cross-validation, where we
compare the mean of the smoothed curves to the observed points of the
left out subject. We then form mean function, covariance function and
nugget effect estimates. The final curves are obtained by going back to
each subject and kriging the curves using the parameter estimates. Such
an approach attempts to better utilize information across subjects to
help with curve construction. This differs slightly from the smoothing
methods found in the PACE package in \citet{matlab2013}, as we
do not utilize a PCA and we construct curves through multiple
refinements as opposed to using scatter plot smoothers. Further details
are outlined in \citet{reimherr2013}.

We present our methods in two sections. The first considers a
functional response with one univariate quantitative covariate and
provides an easier \hyperref[sec1]{Introduction} to our methods. The second section
generalizes the first so that our methods are applicable to a wider
range of settings. While we introduce all technical notation below,
precise mathematical assumptions, theorems and proofs can be found in
Appendices \ref{sasymptotics} and \ref{sproofs}.

\subsection*{Single predictor}
We begin by presenting our procedure for the simpler setting when
\[
Y_n(t) = \beta(t) X_n + \varepsilon_n(t)
\]
for $n=1,\dots,N$, where $X_n$ and $Y_n(t)$ are assumed to be
centered. This will provide a simpler format for introducing the more
technical aspects of FDA tools. We assume that $\{X_n\}$ and $\{
\varepsilon
_n\}$ are two i.i.d. sequences and independent of each other. Without loss
of generality, we will also assume that $t \in[0,1]$. One could assume
that $t$ takes values in any closed interval, but the results will
remain the same.

The slope function $\beta(t)$ we will assume to be square integrable:
\[
\| \beta\|^2 = \int\beta(t)^2 \,dt < \infty.
\]
We will further assume that $\varepsilon_n(t)$ is square integrable
almost surely:
\[
\| \varepsilon_n \|^2 = \int\varepsilon_n(t)^2
\,dt < \infty\qquad\mbox{with probability }1.
\]
Therefore, $\beta$ and $\varepsilon_n$ will take values from the Hilbert
space $L^2[0,1]$ (with probability 1) equipped with the inner product
\[
\langle x, y \rangle= \int x(t) y(t) \,dt.
\]
This implies that $Y_n$ also takes values from $L^2[0,1]$ almost
surely. Throughout, when writing $\| \cdot\|$ of a function, we will
mean this to be the $L^2[0,1]$ norm. We will also assume that
\[
\mathrm{E}\bigl[X_n^2\bigr] < \infty\quad\mbox{and}
\quad\mathrm{E}\| \varepsilon_n\|^2 < \infty.
\]
This will imply [\citet{bosq2000}] that $\varepsilon_n$ will have a
covariance function $C_\varepsilon(t,s)$ which can be expressed as
\[
C_\varepsilon(t,s) = \mathrm{E}\bigl[\varepsilon_n(t)
\varepsilon_n(s)\bigr] = \sum_{i=1}^\infty
\lambda_i v_i(t) v_i(s).
\]
Note that this is simply an application of the spectral theorem. One
can view $C_\varepsilon(t,s)$ as the kernel of an integral operator
acting on
$L^2[0,1]$, in which case $\{\lambda_i\}$ and $\{v_i\}$ will be the
eigenvalues and eigenfunctions, respectively, of the resulting
operator. The moment assumptions will imply that
\[
\sum_{i=1}^\infty\lambda_i <
\infty.
\]
In functional principal component analysis, the $v_i$ are the
functional principal components, while $\lambda_i$ will correspond to
the variability explained by the component. A functional PCA-based
approach would involve choosing a small number of $v_i$ and projecting
the $Y_n$ onto them. This reduces the infinite-dimensional problem
involving $Y_n$ into a multidimensional problem involving the scores
$\langle Y_n,v_i \rangle$. Unfortunately, as we will note in
Section~\ref{sapp}, this can induce stability problems in the resulting
inferential tools.

We can use this assumed structure to construct a test statistic to
determine if $\beta$~is the zero function by examining how much the
inclusion of the covariate reduces the sum of squared norms of the $Y_n$:
\[
\Lambda= \sum_{n=1}^N \| Y_n
\|^2 - \sum_{n=1}^N \|
Y_n - \hat\beta X_n \|^2,
\]
where $\hat\beta$ is the pointwise least squares estimator
\[
\hat\beta(t) = \frac{\sum_{n=1}^N Y_n(t) X_n}{\sum_{n=1}^N X_n^2}.
\]
%
In this simple scenario, the procedure is equivalent to taking $N\|\hat
\beta\|^2$ as the test statistic; however, for testing the nullity of
multiple covariates, $\Lambda$ has a more natural generalization as we
will see in the next section. By Theorem \ref{ttest} in Section~\ref
{sasymptotics}, if $\beta= 0$, then as $N\to\infty$,
\[
\Lambda\stackrel{\mathcal{D}} {\to}\sum_{i=1}^\infty
\lambda_i \chi_i^2(1),
\]
where $\chi_i^2(1)$ are i.i.d. chi-squared 1 random variables. Since the
$\{\lambda_i\}$ are summable, the above will be $O_P(1)$. If $\beta
\neq0$, then
\[
\Lambda= N \mathrm{E}\bigl[X_1^2\bigr] \| \beta
\|^2 + O_P(1)
\]
as $N\to\infty$ by Theorem \ref{ttest}. Such a procedure does not
have the stability problems inherent in PCA techniques and avoids
having to choose the number of components. A minor difficulty arises in
working with the limiting distribution under the null which does not
have a closed-form expression. However, under our assumptions, the
weights $\{\lambda_i\}$ are summable and, in practice, typically
decrease extremely quickly. Thus, one may obtain $p$-values by
considering the distribution of
\[
\sum_{i=1}^I \lambda_i
\chi_i^2(1)
\]
for some large value of $I$. As long as $I$ is reasonably large, the
procedure will be robust against the choice. One can estimate the $\{
\lambda_i\}$ by using the eigenvalues, $\{\hat\lambda_i\}$, of the
empirical covariance function:
\[
\hat C_\varepsilon(t,s) = \frac{1}{N-1} \sum
\bigl(Y_n(t) - \hat\beta(t) X_n\bigr)
\bigl(Y_n(s) - \hat\beta(s) X_n\bigr).
\]
By Theorem \ref{teigen}, the $\{\hat\lambda_i\}$ will be close to
the $\{\lambda_i\}$, uniformly over $i$, for large $N$. Thus, we can
use the estimated eigenvalues to compute $p$-values. While there are
several methods for approximating the distribution of weighted sums of
chi-squares, we have found the method of \citet{imhof1961} to work
very well even for extremely small $p$-values which are required in
genome wide association studies. Further details and comparisons can be
found in \citet{duchesnemicheaux1990}.


\subsection*{Multiple predictors}
In order to test the nullity of multiple predictors, interaction terms
or factor variables all while correcting for other covariates, one
requires a more general testing procedure than found in the previous
section. We now examine the larger model
\begin{eqnarray*}
Y_n(t) & =& \alpha(t) + \sum_{j=1}^J
\beta_{1,j}(t) X_{1,j;n} + \sum_{k=1}^K
\beta_{2,k}(t)X_{2,k;n} + \varepsilon_n(t)
\\
& =& \alpha(t) + \mathbf{X}_{1,n}^T \bolds{
\beta}_1(t) + \mathbf{X}_{2,n}^T \bolds{
\beta}_2(t)+ \varepsilon_n(t). 
\end{eqnarray*}
We assume that $\alpha$, $\{\beta_{1,j}\}$ and $\{\beta_{2,k}\}$ all
take values in $L^2[0,1]$ and that $ \varepsilon_n$ takes values in
$L^2[0,1]$ almost surely. Define the larger vector $\mathbf{X}_n^T = (1,
\mathbf{X}_{1,n}^T, \mathbf{X}_{2,n}^T)$ and assume that $\mathrm
{E}[ \mathbf{X}_n^T \mathbf{X}_n] = \Sigma_X$ exists and has full
rank. Assume
that $\{\mathbf{X}_n\}$ and $\{\varepsilon_n\}$ are two i.i.d. sequences and
independent of each other. As before, assume that $\{\varepsilon_n\}$ are
centered and that $\mathrm{E}\|\varepsilon_n\|^2 < \infty$.

In terms of matrices, the model can be expressed as
\[
\mathbf{Y}(t) = \mathbf{X}_1 \bolds{\beta}_1(t) +
\mathbf{X}_2 \bolds{\beta}_2(t) + \bolds{\varepsilon}(t),
\]
where we group the intercept into the first matrix of covariates. We
abuse notation slightly as $\mathbf{X}_1$ also refers to first
observation, but, given the context, it will always be clear what we mean.
We define the corresponding least squares estimators
\[
\hat{ \bolds{\beta}}_1(t) = \bigl(\mathbf{X}_1^{T}
\mathbf{X}_1\bigr)^{-1} \mathbf{X}_1^{T}
\mathbf{Y}(t) \quad\mbox{and} \quad\hat{ \bolds{\beta}}(t) = \bigl(
\mathbf{X}^{T} \mathbf{X}\bigr)^{-1} \mathbf{X}^{T}
\mathbf{Y}(t),
\]
where
\[
\mathbf{X}= \pmatrix{\mathbf{X}_1 & \mathbf{X}_2}.
\]
We define the more general version of our test statistic as
\[
\Lambda= \sum_{n=1}^N\bigl( \bigl\|
Y_n - \mathbf{X}_{1,n}^T \hat{\bolds{
\beta}}_1\bigr\| ^2 - \| Y_n -
\mathbf{X}_n \hat{\bolds{\beta}}\|^2\bigr).
\]
By Theorem \ref{tmulti} in Appendix~\ref{sasymptotics}, under the
hypothesis that all of the $\bolds{\beta}_2$ coordinates are zero functions,
we have that, as $N\to\infty$,
\[
\Lambda\stackrel{\mathcal{D}} {\to} \sum_{i=1}^\infty
\lambda_i \chi^2_i(K),
\]
where the $\lambda_i$ are as before, but $\chi^2_i(K)$ now have $K$
degrees of freedom corresponding to the $K$ covariates we are testing.
Under the alternative, we have, by Theorem~\ref{tmulti},
\[
\Lambda= N \int\bolds{\beta}_2(t)^T
\Sigma_{X,2\dvtx 1} \bolds{\beta}_2(t) + O_P(1),
\]
where $\Sigma_{X, 2\dvtx 1}$ is the Schur complement
\[
\Sigma_{X,2\dvtx 1} = \Sigma_{X, 22} - \Sigma_{X,21}
\Sigma_{X,11}^{-1} \Sigma_{X,12}.
\]
The matrix $\Sigma_{X,2\dvtx 1}$ appears since we must take into account
how dependent the second set of covariates are on the first. The
empirical eigenvalues (defined as before) $\{\hat\lambda_i\}$ can
again be used for computing $p$-values by Theorem \ref{tmasym}.


\section{A unified framework}\label{sframe}
To better understand the difference between our approach and an FDA
approach based on principal components [\citet{cardotfms2003,kokoszkamaslovasz2008}], we provide a more general
testing framework that includes both.

Consider the single predictor case. In a traditional FDA approach, one
would usually do the following. Using the observations $\{Y_n\}$, form
the estimated eigenvalues and eigenfunctions $\{\hat\lambda_i, \hat
v_i\}$ and assume the $\{X_n\}$ have been standardized such that
\[
N^{-1} \sum_{n=1}^N
X_i = 0 \quad\mbox{and} \quad N^{-1} \sum
_{n=1}^N X_n^2 = 1.
\]
Computing the eigenelements is easily done using the FDA package in
\texttt{R} or \mbox{\texttt{Matlab}}; see \citet{ramsaysilverman2005} for more
details. Then form the test statistic as
\[
\Lambda_2 = \sum_{i=1}^I
\frac{N^{-1} (\sum_{n=1}^N X_n \langle
Y_n, \hat v_i \rangle)^2 }{\hat\lambda_i}.
\]
In other words, one computes the covariance between the covariate and
each PC, then pools the results after standardizing by $\hat\lambda
_j$. We can also express
\[
\Lambda_2 = \sum_{i=1}^I
\frac{N^{-1} (\sum_{n=1}^N X_n \langle
Y_n, \hat v_i \rangle)^2 }{\hat\lambda_j} = \sum_{i=1}^I
\frac{ N \langle\hat\beta, \hat v_i \rangle
^2}{\hat\lambda_i}.
\]
When $\beta= 0$, $\Lambda_2$ is asymptotically $\chi^2(I)$.

Conversely, in this simpler scenario, our $\Lambda$ test statistic becomes
\[
\Lambda= 
N \| \hat\beta\|^2.
\]
By Parceval's identity, we have
\[
N \| \hat\beta\|^2 = N \sum_{i=1}^{N}
\langle\hat\beta, \hat v_i \rangle^2.
\]
So the difference between the two approaches in this simple case can be
thought of as a difference in weighting schemes. For the more general
case, it helps to think in terms of explained variability. Define
\[
\Lambda({ \mathbf w}_N) = N \sum_{i=1}^{N}
w_N(i) \hat R_i^2,
\]
where $\hat R_i^2$ is the proportion of variance in the $i$th PC
explained by the covariates:
\[
\hat R_i^2 = \frac{\sum_{n=1}^N( \langle Y_n, \hat v_i \rangle-
\langle Y_n - X_n \hat\beta, \hat v_i \rangle)^2}{ \sum_{n=1}^N
\langle Y_n, \hat v_i \rangle^2 } = \frac{\langle\hat\beta, \hat v_i
\rangle^2}{ \hat\lambda_i}.
\]
Then $\Lambda_2$ corresponds to taking
\[
w_N(i) = 1_{1 \leq i \leq I},
\]
while $\Lambda$ corresponds to taking
\[
w_N(i) = \hat\lambda_i.
\]
Put in words, each scheme assigns different weights to the projections.
In the traditional case, all (standardized) projections are given equal
weights, while in our approach each (standardized) projection is
weighted by the corresponding amount of variability it explains. Both
weighting schemes arise naturally, but there might be other meaningful
choices of ${\mathbf w}_N$ as well.

Choosing an ``optimal'' weighting scheme is complicated by the fact that
the PCs must be estimated and it is well known that eigenfunction
estimates can be very noisy as one moves beyond the first few PCs.
Furthermore, while $\Lambda$ can be expressed using PCA for comparison
purposes, it does not directly depend on it, which avoids potential
stability problems when dealing with PCs corresponding to small eigenvalues.


\section{Simulation study} \label{ssims}
We carry out a small-scale simulation study to analyze the power of our
procedure as compared to other methods. We generate data using the model
\[
Y_n(t) = \beta(t) X_n + \varepsilon_n(t).
\]
The $X_n$ represent a common SNP with minor allele frequency $0.5$ and
are taken to be i.i.d. binomial random variables with success parameter
$1/2$ and trial parameter $2$, centered by their mean (in this case 1).
The $\varepsilon_n$ are generated as stationary, isotropic, Gaussian
processes with mean zero and covariance coming from the M\'atern class
with parameters $(0,1,0,1/4,5/2)$, representing mean, variance, nugget,
scale and $\nu$, respectively, where $\nu$ controls the smoothness of
the process. In what follows, similar results are
obtained if the $\varepsilon_n$ are replaced with rougher M\'atern
processes or Brownian motion (tables available upon request). Such a
process will have sample paths which are one time continuously
differentiable. The covariance function can be expressed explicitly as,
for $t,s \in[0,1]$ and $d = | t - s|$,
\[
\E\bigl[\varepsilon_n(t) \varepsilon_n(s) \bigr]= C(d)
= \biggl( 1 + \frac{\sqrt{5}d}{1/4} + \frac{5 d^2}{3/16} \biggr) \exp
\biggl( -
\frac{\sqrt{5}d}{1/4} \biggr).
\]
%
%
\begin{figure}[b]
\includegraphics{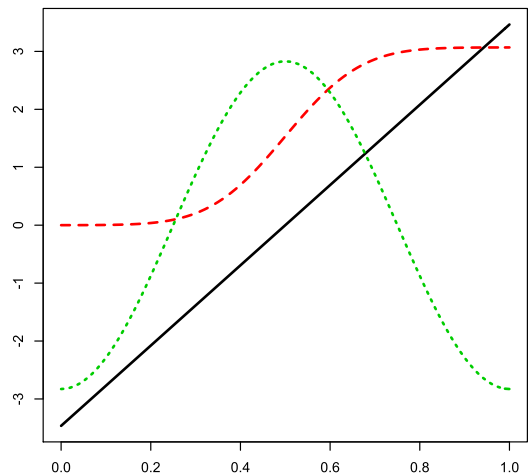}
\caption{Plots of the three $\beta$ functions used under the
alternative hypothesis.}
\label{fbetaplots}
\end{figure}
We take $N=200$, use $1000$ repetitions in all cases, and let $M =
5,10,20,50$, where $M$ is the number of points sampled per curve. We
assume that the points are always sampled on an even grid on the
$[1/M,1]$ interval and the curves are\vadjust{\goodbreak} reconstructed using the same
approach mentioned at the beginning of Section~\ref{smeth}. We
compare three different slope functions:
\begin{enumerate}[3.]
\item Linear Function: $\beta(t) = 0.18 \times2(t - 1/2)
/0.5773$, 
\item Normal CDF: $\beta(t) = 0.18 \times\Phi(7.5(t -
1/2))/0.6517 $, 
\item Sinusoidal: $\beta(t) = 0.18 \times\sqrt{2}\cos
(2 \pi t)$. 
\end{enumerate}
Notice that the functions are normalized such that the $L^2$ norm of
the function is $0.18$, which was chosen to get a clear comparison of
power between the procedures. We are especially
interested in the second, as we believe its shape to be more reflective
of the types of patterns we expect to see in our asthma data as well as
human growth data in general; over the course of several years children
grow in spurts followed by a leveling off as they get closer to
adulthood. Plots of the above functions are given in Figure~\ref{fbetaplots}.

%
%
\begin{figure}[t]
\includegraphics{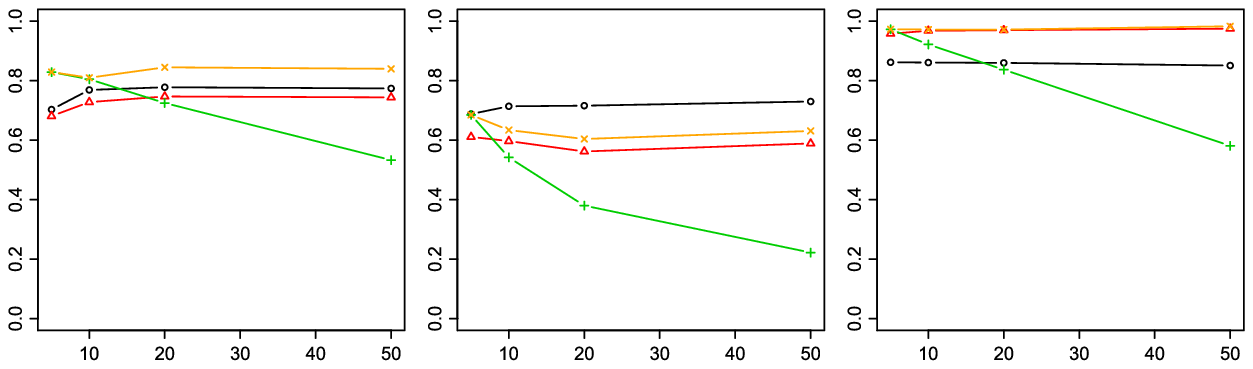}
\caption{Power plots for 4 procedures plotted against the number of
points sample per curve ($M$). Method L2 is $\bolds{\circ}$, PC
is \protect\includegraphics{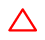}, PC5 is \protect\includegraphics{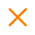}, and MV is
\protect\includegraphics[raise=-1pt]{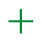}.
The left, middle and right panels correspond to the
linear, normal c.d.f. and sinusoidal signals.}\label{fpower}
\end{figure}

The methods we compare in each scenario are as follows:
\begin{itemize}
\item L2: Our Method ($\bolds{\circ}$),

\item PC: A MANOVA performed with 3, 4, and total 5 PCs,
the $p$-value is then taken as the largest (\includegraphics{692i01.eps}),

\item PC5: A MANOVA performed with 5 PCs (\includegraphics{692i02.eps}),

\item MV: A MANOVA performed on the observed points (\includegraphics{692i03.eps}).
\end{itemize}

In parenthesis we include the plotting symbol for each procedure
plotted in Figure~\ref{fpower}. We include both Methods PC and PC5 to
illustrate the consequences of having to choose the number of PCs.
Method PC is common when trying to determine if one's results are
robust against the number of PCs chosen. Note that Methods L2, PC5 and
MV are well calibrated, while Method 2 is conservative
(tables available upon request, not shown here for brevity).

%
%
\begin{figure}[b]
\includegraphics{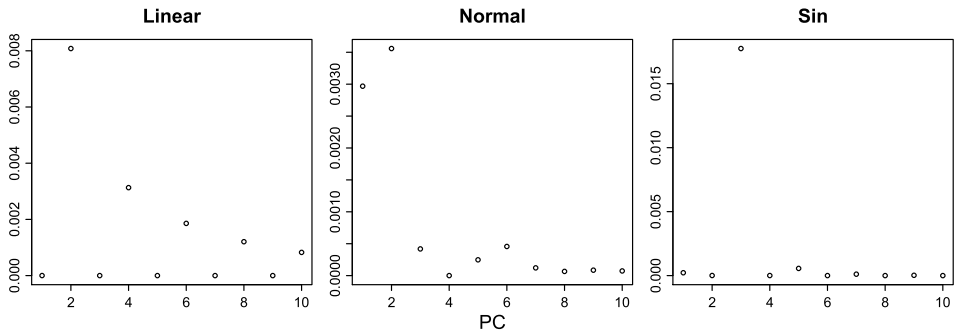}
\caption{$R^2$ plotted against PC number.}
\label{frsq}
\end{figure}

The results are summarized in Figure~\ref{fpower}. There are a few
interesting patterns that become clear upon examination. The first is
that the PC5 method does quite well in all of the settings. However,
the PC method does a bit worse, as expected. This reflects the price of
having to choose the number of PCs. Our procedure's behavior is fairly
consistent across scenarios due to the normalization of the $\beta$
functions. The smoothness of the process seems to help our method, and
in the normal c.d.f. setting we seem to do quite well, while in the
sinusoidal case our method performs slightly worse. The MV method seems
to not work as well when compared to the FDA methods. It is especially
interesting to see that the multivariate approach does worse as one
adds points. This is due to the high dependence of the points across
time; increasing the number of points increases the dimension, while
the high correlation between points means the overall signal does not
increase much.

We conclude this section with a closer examination of what is driving
the simulation results. Given the exposition in Section~\ref{sframe},
we can gain insight by looking at the $R^2$ for each PC. That is, we
project the $Y_n$ onto each PC and in each case look at what proportion
of the variability is explained by the covariate. In Figure~\ref{frsq}
we plot the theoretical $R^2$ values against the PCs.

As we can see, the patterns provide a nice reflection of the power
results. For the normal setting, the $R^2$ decreases with the PC
number, while in the sinusoidal setting one gets a peak at PC number 3
(because here $\beta$ closely matches the 3rd PC). The linear setting
has a very interesting alternating pattern (due to the PCs alternating
as even/odd functions). When looking at plots of $\langle\beta, v_j
\rangle^2$ only (not shown here for brevity), one sees similar
patterns, but they are more subtle. It is not until those plots are
scaled by the eigenvalues that you see these very strong patterns. An
important note is that to obtain a similar looking plot empirically,
that is, using empirical $R^2$ values, one needs a rather large sample
size. For $N=200$, one will obtain a very chaotic plot. It is not until
$N$ is over 1000, or better yet 10,000, that one sees the empirical
plots agreeing nicely with the theoretical ones. This is in large part
due to the difficulty in estimating eigenfunctions. Eigenvalues are, in
some sense, much easier to estimate accurately. However, the accuracy
of eigenfunction estimates depends greatly on what the eigenvalues are,
and, in particular, how large and spread out the values are. The
smaller an eigenvalue is, and the closer it is to another eigenvalue,
the harder the corresponding eigenfunction becomes to estimate.

\section{Application} \label{sapp}

The childhood asthma management project, CAMP, is a multi-center,
longitudinal clinical trial designed to better understand the long-term
impact of several treatments for mild to moderate asthma
[The Childhood Asthma Management Program
Research Group (\citeyear{camp1999,camp2000})]. Subjects, ages 5--12, with asthma were
selected, randomly assigned a particular treatment and monitored for
several years. Our data consists of 540 Caucasian subjects monitored
for 4 years, each of whom made 16 clinical visits. Genome-wide SNP
data and phenotype information were downloaded from dbGaP
(\url{http://www.ncbi.nlm.nih.gov/gap}) study accession phs000166.v2.p1. Each
subject's first three visits are 1--2 weeks apart and occur prior to
treatment; the second two visits are 2 months apart, while the
remaining 11 visits are around 4--5 months apart. While a large number
of measurements are taken, we focus on the log of \textit{forced
expiratory volume in one second} (FEV1), that is, the total volume of
air a subject can force out of their lungs in one second. Each subject
is given one of two treatments, Budesonide and Nedocromil, or assigned
to the Placebo group. Each treatment was assigned to approximately $30\%
$ of the subjects, with the remaining $40\%$ receiving the placebo.
Trajectories were assembled using a smoothing approach based on
B-splines, the details of which can be found in \citet{reimherr2013}.

%
%
\begin{figure}[t]
\includegraphics{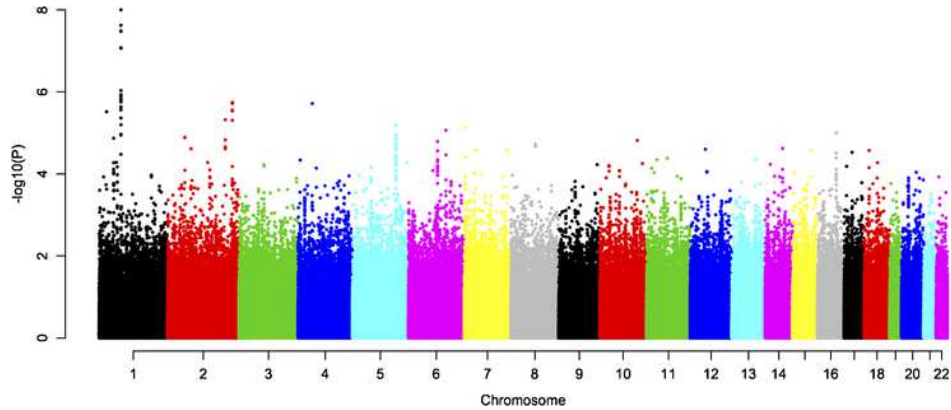}
\caption{Manhattan plot of GWAS $p$-values across chromosomes.}
\label{fman}
\end{figure}

%
%
\begin{figure}[b]
\includegraphics{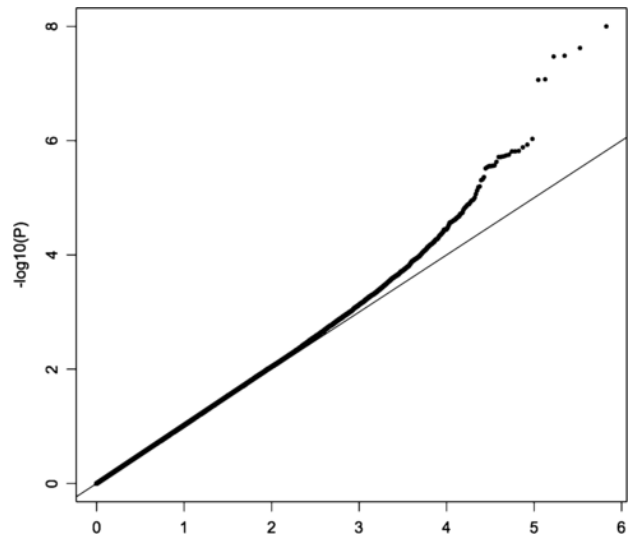}
\caption{
QQ-plot comparing GWAS $p$-values ($y$-axis, observed) to uniform
distribution ($x$-axis, expected) on $\log_{10}$ scale.}\label{fqq}
\end{figure}

\begin{figure}
\includegraphics{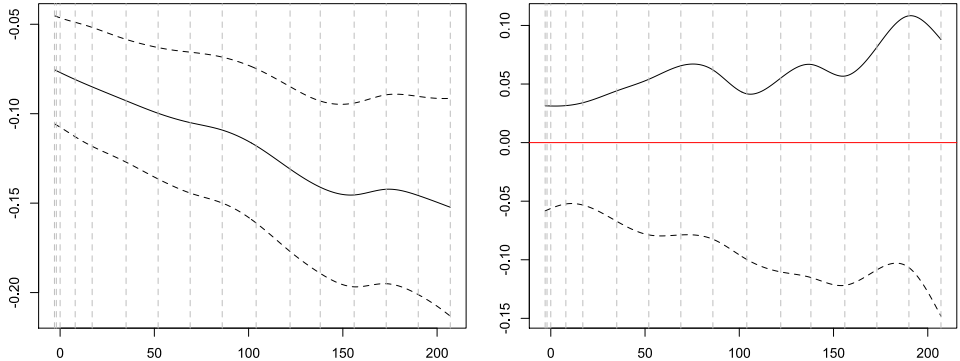}
\caption{The left panel is a plot of the SNP ($\mathit{rs}12734254$) effect
with pointwise confidence intervals included as dashed lines.
The right panel is a plot of the SNP by treatment interaction with
Placebo set as the baseline (in red), Budesonide is given by the solid
line, and Nedocromil as the dashed. We include grey vertical lines
indicating the time of the visits.}
\label{fint}
\end{figure}

For each subject we have approximately six hundred and seventy thousand
SNPs genotyped after filtering out those with minor allele frequencies
below $5\%$. We used our procedure to test for an association between
FEV1 and each SNP, while correcting for age, gender and treatment. The
model is given by (at each time point)
\[
\log(\mathrm{FEV}1) \sim \mbox{age} + \mbox{gender} + \mbox{treatment} + \mbox{SNP},
\]
where we take age to be the age of the patient at the beginning of the
study. A~Manhattan plot summarizing all the $p$-values across
chromosomes is given in Figure~\ref{fman} and a QQ-plot on the
$\log_{10}$ scale is given in Figure~\ref{fqq}. Examining the
QQ-plot, we can see that the procedure is well calibrated, as the
$p$-values smaller than 0.001 fall directly on the 45 degree line, while
there are also some very small \mbox{$p$-}values indicating some genetic
associations. We found one SNP with a $p$-value of $1.016 \times
10^{-8}$: $\mathit{rs}12734254$ in gene ST6GALNAC5 on chromosome 1 (significant
at the $5\%$ significance level with a Bonferroni correction). A plot
of the estimated coefficient function is given in Figure~\ref{fint}.
As we can see, presence of the minor allele (frequency $41.63 \%$) is
associated with a decrease in lung function which worsens with time.
The magnitude of the $p$-value remained the same for
different levels of smoothing in the preprocessing step, thus, the
finding is fairly robust against the initial level of smoothing.

%
%

Next, we tested if that SNP also had an interaction effect with
treatment; we found a $p$-value of 0.00284. Since we used significance
of the main effects as type of screening before testing for an
interaction (which resulted in one test), we do not need to make a
correction for multiple testing and, thus, the $p$-value gives strong
evidence that there is a SNP by treatment interaction. In Figure~\ref
{fint} we plot the SNP by treatment interaction effect so that we can
better understand the nature of the interaction (with the placebo group
set as the baseline). As we can see, the interaction seems to be driven
by the difference between the Budesonide group and the Nedocromil
group. The Nedocromil group has a stronger decreasing trend, while the
Budesonide effect is actually positive. Thus, the presence of the minor
allele suggests that Budesonide might in fact have a positive impact on
lung function or that at least Budesonide can counter the effects of
the deleterious SNP.

To illustrate the power gained by using the FDA approach, we compared
it to a univariate approach. If instead of using our approach we take
the difference between the endpoints and perform the same type of
association analysis (but now with a scalar response), the $p$-value for
our SNP effect becomes $0.002$ and the \mbox{$p$-}value for the interaction
test becomes $0.135$. Since this is our top SNP, we would expect the
$p$-value for the scalar approach to be smaller, but the drop in
$p$-value is still quite large. More generally, one does not need an FDA
approach to exploit the temporal structure in longitudinal data, but it
does give a natural framework in which to do it. The potential power
gains made by meaningfully pooling across coordinates should be a
serious consideration in any longitudinal approach.

Finally, we note that when using an FDA approach based on PCA we ran
into some interesting stability issues. In particular, we would on
occasion see some very large signals coming from a SNP where the rare
allele was only present in a few individuals, making the result
unreliable and suspicious. Upon closer examination, it seemed as though
the PCs corresponding to smaller eigenvalues were the culprit. This was
likely due to the inherent noise in those eigenfunctions which may have
been driven by a small number of individuals. Typically, this would not
obviously be problematic, but when carrying out a GWAS with hundreds of
thousands of SNPs it is possible to come upon a SNP present only in
those individuals which are driving the smaller PCs.

\section{Discussion} \label{sdisc}
We have presented a new method based on functional data analysis for
detecting genetic associations with longitudinal variables. FDA methods
allow for a flexible framework that exploits the temporal structure of
the data which can result in increased power. Two
primary advantages of our particular approach over a PCA-based test is
that one does not need to choose the number of principal components
(while still maintaining excellent power) and the interpretation of the
results is not tied to the interpretability of the shapes of the PCs,
which can be especially challenging for the nonleading eigenfunctions.
Furthermore, we have found our procedure to be substantially more
stable than a PC-based method. This became crucial when conducting
hundreds of thousands of tests. We showed how the smoothness of the
trajectories and underlying parameter functions determine the power of
our procedure, as well as traditional PCA-based methods, as compared to
a multivariate approach. In particular, our simulations reinforced the
current FDA paradigm: the smoother the objects, the greater the
advantage of FDA procedures.

We applied our methodology to data coming from the childhood asthma
management study (CAMP). We showed how an association test can be
carried out which measures the effect of a SNP on a functional object,
while correcting for other covariates. We then followed that test up
with an interaction test. Interestingly, we found a mildly significant
SNP effect with a significant drug by SNP interaction effect. Such
results are of great interest due to the impact they can have on
choosing treatment courses for patients.

A useful area for improvement would be to develop a
method for estimating/incorporating the interpolation error from the
preprocessing step. In high frequency settings such an error is
typically very small, but in longitudinal settings where one has a
relatively small number of observations per subject, the error can be
significant. Indeed, for a very small number of points per curve (2--4)
the interpolation error would be so large that classical multivariate
methods or methods based on pooled nonparametric smoothers might be
more appropriate. Accounting for this error would hopefully help with
power and parameter estimation.

We would be excited to see more FDA methodology developed that does not
directly depend on PCA. While we have presented a method based on $L^2$
norms, it would be interesting to see how one could account for
processes whose covariance changes substantially over time. We
mentioned the idea of standardizing the functions to have unit variance
at every time point. However, it is still unclear how one could take
into account the covariance between time points. Indeed, it might be
desirable to give less weight to temporal regions with high dependence,
since they, in some sense, carry less information than regions with
lower dependence.

We believe FDA to be a very interesting and promising method for
exploiting the temporal structure in longitudinal studies. The problems
facing geneticists are quite challenging due to the sheer abundance of
noise inherent in their data. To help overcome this noise, researchers
are constantly finding new ways of exploiting information and
structure. Here we have shown how an FDA method can exploit a~temporal
structure to achieve better power. Furthermore, the results are still
interpretable as we are, in essence, focusing on large scale patterns.
Thus, as we saw in our application, we were able to deliver powerful
results with interpretable conclusions. While the FDA toolbox is
expanding rapidly, very little has been done which tailors FDA
techniques for genetic studies. We hope that we have taken a~meaningful
step in this direction.

\begin{appendix}
\section{Asymptotic results} \label{sasymptotics}
In this section we provide the explicit underlying assumptions and
asymptotic results which justify the behavior of our procedure under
the null and alternative hypotheses.

\subsection*{Single predictor}
%
%
\begin{assumption}\label{a1}
Assume we have the following relationship:
\[
Y(t) = \beta(t) X + \varepsilon(t),
\]
where $X$ and $\varepsilon$ are random and take values in $\mathbb
{R}$ and $L^2[0,1]$, respectively. Assume that $\beta\in L^2[0,1]$ and
that the following moment conditions hold:
\[
EX = 0,\qquad \E\bigl[\varepsilon(t)\bigr] = 0, \qquad EX^2 < \infty
\quad\mbox{and} \quad \E\|\varepsilon\|^2 < \infty.
\]
Finally, assume that $(X_1, Y_1), \ldots, (X_N, Y_N)$ are i.i.d. copies of $(X,Y)$.
\end{assumption}

%
%
\begin{theorem}\label{tbeta}
If Assumptions \ref{a1} holds, then
\[
\sqrt{N}(\hat{\beta} - \beta) \stackrel{\mathcal{D}} {\to} Z,
\]
where $Z$ is a mean zero Gaussian process in $L^2[0,1]$ with covariance operator
\[
\frac{\E[\varepsilon\otimes\varepsilon]}{\E[X^2]}.
\]
\end{theorem}

%
%
\begin{theorem} \label{ttest}
If Assumptions \ref{a1} holds and $\beta= 0$, then
\[
\Lambda\stackrel{\mathcal{D}} {\to}\sum_{i=1}^\infty
\lambda_i \chi_i^2(1),
\]
where $\{\lambda_i\}$ are the eigenvalues of the covariance operator
of $\varepsilon$ and $\{\chi_i^2(1)\}$ are i.i.d. chi-squared 1 random
variables.
If $\beta\neq0$ (in an $L^2$ sense), then
\[
\Lambda= N\mathrm{E}\bigl[X^2\bigr] \| \beta\|^2 +
O_P(1).
\]
\end{theorem}

%
%
\begin{theorem}\label{teigen}
If Assumption \ref{a1} holds, then
\[
\hat C_\varepsilon:= \frac{1}{N-1} \sum_{n=1}^N
(Y_n - X_n \hat\beta) \otimes(Y_n -
X_n \hat\beta) \stackrel{P} {\to} C_\varepsilon,
\]
where convergence occurs in the space of Hilbert--Schmidt operators
and, consequently,
\[
\sup_{1 \leq i < \infty}| \hat\lambda_i - \lambda_i|
\leq\|\hat C_\varepsilon- C_\varepsilon\| \stackrel{P} {\to}0,
\]
where $\{\hat\lambda_i\}$ are the eigenvalues of $\hat C_\varepsilon
$. If
in addition $\E\|\varepsilon_n\|^4<\infty$, then
\[
\sqrt{N}(\hat C_\varepsilon- C_\varepsilon) \stackrel{\mathcal{D}} {\to}
\mathcal{N}(0, \Gamma),
\]
where $\Gamma= \E[(\varepsilon_1 \otimes\varepsilon_1 -
C_\varepsilon) \otimes(\varepsilon_1 \otimes\varepsilon_1 -
C_\varepsilon)]$. Consequently, one has that
\[
\sup_{1 \leq i < \infty}| \hat\lambda_i - \lambda_i|
\leq\|\hat C_\varepsilon- C_\varepsilon\| = O_P
\bigl(N^{-1/2}\bigr).
\]
\end{theorem}

\subsection*{Multiple predictors}
%
%
\begin{assumption}\label{a3}
Assume we have the following relationship:
\begin{eqnarray*}
Y_n(t) & =& \alpha(t) + \sum_{j=1}^J
\beta_{1,j}(t) X_{1,j;n} + \sum_{k=1}^K
\beta_{2,k}(t)X_{2,k;n} + \varepsilon_n(t)
\\
& =& \alpha(t) + \mathbf{X}_{1,n}^T \bolds{
\beta}_1(t) + \mathbf{X}_{2,n}^T \bolds{
\beta}_2(t)+ \varepsilon_n(t),
\end{eqnarray*}
where $J$ and $K$ are fixed integers and $\mathbf{X}_{1,n}, \mathbf
{X}_{2,n}$ and $\varepsilon$ are random and take values in $\mathbb
{R}^J$, $\mathbb{R}^K$ and $L^2[0,1]$, respectively. Assume that $\{
\beta_{1,j}\}$ and $\{\beta_{2,k}\}$ are elements of $ L^2[0,1]$ and
that $\E\|\varepsilon\|^2 < \infty$. Let $\mathbf{X}_n^T = (1,
\mathbf
{X}_{1,n}^T, \mathbf{X}_{2,n}^T)$ and assume that $\mathrm{E}[\mathbf
{X}_n^T\mathbf{X}_n] = \Sigma_X$ exists and has full rank. Last,
assume that $\{\mathbf{X}_1, \ldots, \mathbf{X}_N\}$ and $\{
\varepsilon_1, \ldots, \varepsilon_N \}$ are two i.i.d. sequences,
independent of each other, and that $\mathrm{E}[\varepsilon_n(t)] = 0$.
\end{assumption}

%
%
\begin{theorem}\label{tmulti}
If Assumptions \ref{a3} holds and $ \bolds{\beta}_2 =0$, then
\[
\Lambda\stackrel{\mathcal{D}} {\to} \sum_{i=1}^\infty
\lambda_i \chi_i^2(K),
\]
where $\{\lambda_i\}$ are the eigenvalues of the covariance operator
of $\varepsilon$ and $\{\chi_i^2(K)\}$ are i.i.d. chi-squared $K$
random variables. If $\bolds{\beta}_2 \neq0$, then
\[
\Lambda= N \int\bolds{\beta}_2(t)^T
\Sigma_{X,2\dvtx 1} \bolds{\beta}_2(t) + O_P(1),
\]
where $\Sigma_{X, 2\dvtx 1}$ is the Schur complement
\[
\Sigma_{X,2\dvtx 1} = \Sigma_{X, 22} - \Sigma_{X,21}
\Sigma_{X,11}^{-1} \Sigma_{X,12}.
\]
\end{theorem}

%
%
\begin{theorem}\label{tmasym}
If Assumptions \ref{a3} holds, then
\[
\sqrt{N}(\hat{\bolds{\beta}}- \bolds{\beta}) \stackrel{\mathcal{D}}
{\to}
\mathcal{N}\bigl(0, \Sigma^{-1} C_{\varepsilon}\bigr),
\]
where convergence occurs with respect to the product space
$(L^2[0,1])^N$. Furthermore, we have
\[
\hat C_\varepsilon:= \frac{1}{N-1-J-K} \sum_{n=1}^N
(Y_n - \mathbf{X}_n \hat{\bolds{\beta}})
\otimes(Y_n - \mathbf{X}_n \hat{\bolds{\beta}})
\stackrel{P} {\to} C_\varepsilon,
\]
where convergence occurs in the space of Hilbert--Schmidt operators
and, consequently,
\[
\sup_{1 \leq i < \infty}| \hat\lambda_i - \lambda_i|
\leq\|\hat C_\varepsilon- C_\varepsilon\| \stackrel{P} {\to}0,
\]
where $\{\hat\lambda_i\}$ are the eigenvalues of $\hat C_\varepsilon
$. If
in addition $\E\|\varepsilon_n\|^4<\infty$, then
\[
\sqrt{N}(\hat C_\varepsilon- C_\varepsilon) \stackrel{\mathcal{D}} {\to}
\mathcal{N}(0, \Gamma),
\]
where $\Gamma= \E[(\varepsilon_1 \otimes\varepsilon_1 -
C_\varepsilon) \otimes(\varepsilon_1 \otimes\varepsilon_1 -
C_\varepsilon)]$. Consequently, one has
\[
\sup_{1 \leq i < \infty}| \hat\lambda_i - \lambda_i|
\leq\|\hat C_\varepsilon- C_\varepsilon\| = O_P
\bigl(N^{-1/2}\bigr).
\]
\end{theorem}


\section{Proofs} \label{sproofs}
Since the simpler single predictor scenario is a special case of the
more general setting, we will only prove the more general theorems.

\begin{pf*}{Proof of Theorem \ref{tmulti}}
In matrix form we can express the model as
\[
\mathbf{Y}(t) = \mathbf{X}_1\bolds{\beta}_1(t) +
\mathbf{X}_2 \bolds{\beta}_2(t) + \bolds{\varepsilon} (t)=
\mathbf{X}\bolds{\beta}(t) + \bolds{\varepsilon}(t).
\]
We use the least squares estimators
\[
\hat{\bolds{\beta}}(t) = \bigl(\mathbf{X}^T \mathbf{X}
\bigr)^{-1} \mathbf{X}^T \mathbf{Y}(t) \qquad\hat{\bolds{
\beta}}_1(t) = \bigl(\mathbf{X}_1^T
\mathbf{X}_1\bigr)^{-1} \mathbf{X}_1^T
\mathbf{Y}(t).
\]
We define our test statistic as
\[
\Lambda_2 = \sum_{n=1}^N \bigl\|
Y_n(t) - \mathbf{X}_{1,n}^T \hat{\bolds{
\beta}}_1(t) \bigr\|^2 - \sum_{n=1}^N
\bigl\| Y_n(t) - \mathbf{X}_n^T \hat{\bolds{
\beta}}(t) \bigr\|^2.
\]
The sum of squared residuals based on $\mathbf{X}$ can now be
expressed as
\begin{eqnarray*}
&& \bigl(\mathbf{Y}(t) - \mathbf{X}\hat{\bolds{\beta}}(t)\bigr)^T
\bigl(\mathbf{Y}(t) - \mathbf{X}\hat{\bolds{\beta}}(t)\bigr)
\\
&&\qquad = \bigl( \bolds{\varepsilon}(t) - \mathbf{X}\bigl(\mathbf{X}^T
\mathbf{X}\bigr)^{-1} \mathbf{X}^T \bolds{\varepsilon}(t)
\bigr)^T \bigl( \bolds{\varepsilon}(t) - \mathbf{X}\bigl(
\mathbf{X}^T \mathbf{X}\bigr)^{-1} \mathbf{X}^T
\bolds{\varepsilon}(t)\bigr)
\\
&&\qquad = \bolds{\varepsilon}(t)^T \bolds{\varepsilon}(t) - \bolds{
\varepsilon}(t)^T \mathbf{X}\bigl(\mathbf{X}^T \mathbf{X}
\bigr)^{-1} \mathbf{X}\bolds{\varepsilon}(t).
\end{eqnarray*}
The residuals using only $\mathbf{X}_1$ can be expressed as
\begin{eqnarray*}
&& \mathbf{Y}(t) - \mathbf{X}_1 \hat{\bolds{\beta}}_1(t)
\\
&&\qquad = \mathbf{X}\bolds{\beta}(t) + \bolds{\varepsilon}(t) - \mathbf
{X}_1\bigl(\mathbf{X}_1^T
\mathbf{X}_1\bigr)^{-1} \mathbf{X}_1^T
\mathbf{Y}(t)
\\
&&\qquad = \mathbf{X}\bolds{\beta}(t) + \bolds{\varepsilon}(t) - \mathbf
{X}_1\bigl(\mathbf{X}_1^T
\mathbf{X}_1\bigr)^{-1} \mathbf{X}_1^T
\mathbf{X}\bolds{\beta}(t) - \mathbf{X}_1\bigl(
\mathbf{X}_1^T \mathbf{X}_1
\bigr)^{-1} \mathbf{X}_1^T\bolds{
\varepsilon}(t)
\\
&&\qquad = \mathbf{X}_2 \bolds{\beta}_2(t) + \bolds{
\varepsilon}(t) - \mathbf{X}_1\bigl(\mathbf{X}_1^T
\mathbf{X}_1\bigr)^{-1} \mathbf{X}_1^T
\mathbf{X}_2 \bolds{\beta}_2(t) - \mathbf{X}_1
\bigl(\mathbf{X}_1^T \mathbf{X}_1
\bigr)^{-1} \mathbf{X}_1^T\bolds{
\varepsilon}(t).
\end{eqnarray*}
So the sum of squared residuals is (after expanding and combining like terms)
\begin{eqnarray*}
&& \bolds{\beta}_2(t)^T \mathbf{X}_2^T
\mathbf{X}_2 \bolds{\beta}_2(t) +\bolds{
\varepsilon}(t)^T \bolds{\varepsilon}(t) - \bolds{
\beta}_2(t)^T \mathbf{X}_2^T
\mathbf{X}_1\bigl(\mathbf{X}_1^T
\mathbf{X}_1\bigr)^{-1} \mathbf{X}_1^T
\mathbf{X}_2 \bolds{\beta}_2(t)
\\
&&\qquad{} - \bolds{\varepsilon}(t)^T \mathbf{X}_1\bigl(
\mathbf{X}_1^T \mathbf{X}_1
\bigr)^{-1} \mathbf{X}_1^T\bolds{
\varepsilon}(t) + 2 \bolds{\beta}^T_2(t)
\mathbf{X}_2^T \bolds{\varepsilon}(t)
\\
&&\qquad{} - 2 \bolds{\beta}_2(t)^T \mathbf{X}_2^T
\mathbf{X}_1\bigl(\mathbf{X}_1^T
\mathbf{X}_1\bigr)^{-1} \mathbf{X}_1^T
\bolds{\varepsilon}(t).
\end{eqnarray*}
By examining the orders of each of the terms, one can verify that for
$\bolds{\beta}_2 \neq0$ we have
\[
\Lambda_2 = N \int\bolds{\beta}_2(t)^T
\Sigma_{X,2\dvtx 1} \bolds{\beta}_2(t) \,dt+O_P(1).
\]
Since $\Sigma_{X}$ has full rank, $\Sigma_{X,2\dvtx 1}$ is positive
definite. Therefore, under the alternative we have
\[
\Lambda_2 \to\infty.
\]
Under the null $\bolds{\beta}_2 = 0$, the sum of squared residuals becomes
\[
\bolds{\varepsilon}(t)^T \bolds{\varepsilon}(t) - \bolds{
\varepsilon}(t)^T \mathbf{X}_1\bigl(\mathbf{X}_1^T
\mathbf{X}_1\bigr)^{-1} \mathbf{X}_1^T
\bolds{\varepsilon}(t).
\]
Therefore, the reduction in the sum of squared residuals by including
$\mathbf{X}_2$ is given by
\begin{eqnarray*}
&& \bolds{\varepsilon}(t)^T \mathbf{X}\bigl(\mathbf{X}^T
\mathbf{X}\bigr)^{-1} \mathbf{X}^T \bolds{\varepsilon}(t) -
\bolds{\varepsilon}(t)^T \mathbf{X}_1\bigl(
\mathbf{X}_1^T \mathbf{X}_1
\bigr)^{-1} \mathbf{X}_1^T \bolds{
\varepsilon}(t)
\\
&&\qquad = \bolds{\varepsilon}(t)^T \bigl[ \mathbf{X}\bigl(
\mathbf{X}^T \mathbf{X}\bigr)^{-1} \mathbf{X}^T
- \mathbf{X}_1\bigl(\mathbf{X}_1^T
\mathbf{X}_1\bigr)^{-1} \mathbf{X}_1^T
\bigr] \bolds{\varepsilon}(t).
\end{eqnarray*}
Using the Hilbert space CLT, we have that
\[
N^{-1/2} \mathbf{X}^T\bolds{\varepsilon}(t) \stackrel{
\mathcal{D}} {\to} \mathbf{Z}(t),
\]
where $\mathbf{Z}(t)$ is a vector of Gaussian processes that can be
expressed as
\[
\mathbf{Z}(t) = \Sigma_X^{1/2} \mathbf{Z}^\prime(t),
\]
where $\mathbf{Z}^\prime(t)$ is a vector of i.i.d. Gaussian processes with
covariance functions $\E[\varepsilon_n(t) \varepsilon_n(s)]$.
Therefore, the reduction in the sum of squared residuals (by the
continuous mapping theorem and Slutsky's lemma) is asymptotically equal
in distribution to
\[
\mathbf{Z}^\prime(t)^T \mathbf{A}\mathbf{Z}^\prime(t),
\]
where
\[
\mathbf{A}= \mathbf{I}- \Sigma_x^{1/2} \pmatrix{
\Sigma_{x,11}^{-1} & 0
\vspace*{3pt}\cr
0 & 0 }\Sigma_x^{1/2}.
\]
Notice that $\mathbf{A}$ is in fact a projection matrix with rank
\[
\operatorname{rank}(\mathbf{A}) = \operatorname{trace}(\mathbf{A}) = (K+J) - J = K.
\]
Therefore, we have that
\[
\mathbf{Z}^\prime(t)^T \mathbf{A}\mathbf{Z}^\prime(t)
\stackrel{\mathcal{D}} { = } \sum_{k=1}^K
Z_k^\prime(t)^2.
\]
This implies that the asymptotic distribution of $\Lambda_2$ is now a
weighted sum of $\chi^2(K)$ random variables.
\end{pf*}

\begin{pf*}{Proof of Theorem \ref{tmasym}}
We start by showing the asymptotic normality of $\hat{\bolds{\beta}}$. Notice
we can express
\[
\sqrt{N}\bigl(\hat{\bolds{\beta}}(t) - \bolds{\beta}(t)\bigr) = \sqrt
{N}\bigl(
\mathbf{X}^T \mathbf{X}\bigr)^{-1} \mathbf{X}^T
\mathbf{Y}(t) = \bigl(N^{-1} \mathbf{X}^T \mathbf{X}
\bigr)^{-1} \bigl(N^{-1/2} \mathbf{X}^T
\mathbf{Y}(t)\bigr).
\]
By the multivariate law of large numbers,
\[
N^{-1} \mathbf{X}^T \mathbf{X}\to\Sigma_X.
\]
By the CLT for Hilbert spaces,
\[
\mathbf{X}^T \mathbf{Y}\stackrel{\mathcal{D}} {\to} \mathcal{N}(0,
\Sigma C).
\]
So by Slutsky's lemma,
\[
\sqrt{N}(\hat{\bolds{\beta}}- \bolds{\beta}) \stackrel{\mathcal{D}}
{\to}
\mathcal{N}\bigl(0, \Sigma^{-1} C\bigr)
\]
as desired.

Next, turning to the estimate of the covariance operator for the error
terms, we have
\begin{eqnarray*}
\hat C_\varepsilon& =& \frac{1}{N-1-J-K} \sum_{n=1}^N(Y_n
- \mathbf{X}_n \hat{\bolds{\beta}}) \otimes(Y_n -
\mathbf{X}_n \hat{\bolds{\beta}})
\\
& =& \frac{1}{N-1-J-K} \sum_{n=1}^N\bigl(
\varepsilon_n - \mathbf{X}_n (\hat{\bolds{\beta}}-
\bolds{\beta}) \bigr) \otimes\bigl(\varepsilon_n -
\mathbf{X}_n (\hat{\bolds{\beta}}- \bolds{\beta}) \bigr).
\end{eqnarray*}
Examining the pieces, we have that
\[
\sum_{n=1}^N \mathbf{X}_n (
\hat{\bolds{\beta}}- \bolds{\beta}) \otimes\varepsilon_n = \sum
_{n=1}^N \mathbf{X}_n (\hat{
\bolds{\beta}}- \bolds{\beta}) \otimes\varepsilon_n =
O_P(1)
\]
by combining the convergence rate of $\hat{\bolds{\beta}}$ and the Hilbert
space CLT. Next we have
\[
\sum_{n=1}^N \mathbf{X}_n (
\hat{\bolds{\beta}}- \bolds{\beta}) \otimes\mathbf{X}_n (\hat{
\bolds{\beta}}- \bolds{\beta}) = O_P(1)
\]
by combining the convergence rate of $\hat{\bolds{\beta}}$ and the
multivariate law of large numbers. Therefore, we can conclude
\[
\hat C_\varepsilon= N^{-1} \sum_{n=1}^N
\varepsilon_n \otimes\varepsilon_n + O_P
\bigl(N^{-1}\bigr).
\]
We then immediately have that
\[
C_\varepsilon\stackrel{P} {\to} C_\varepsilon
\]
by the Hilbert space law of large numbers. Notice that
\[
\|\varepsilon_n \otimes\varepsilon_n\|^2 =
\int\!\!\int\varepsilon_n(t)^2 \varepsilon_n(s)^2
\,dt \,ds = \| \varepsilon_n\|^4.
\]
Therefore, $\mathrm{E}\|\varepsilon_n\|^4 < \infty$ implies that
$\mathrm{E}\|\varepsilon_n \otimes\varepsilon_n\|^2<\infty$ and by
the Hilbert space CLT we can conclude that
\[
\sqrt N ( \hat C_\varepsilon- C_\varepsilon) \to N(0, \Gamma),
\]
where
\[
\Gamma= \mathrm{E}\bigl[( \varepsilon_n \otimes
\varepsilon_n - C_\varepsilon) \otimes( \varepsilon_n
\otimes\varepsilon_n - C_\varepsilon)\bigr].
\]
To obtain the final claim, we apply Corollary 4.5 on page 252 of \citet{gohberg2003}, which gives
\[
|\hat\lambda_i - \lambda_i| \leq\| \hat
C_{\varepsilon} - C_{\varepsilon}\|.
\]\upqed
\end{pf*}
\end{appendix}



%

%
\endbibitem

\printaddresses

\end{document}